\documentclass[12pt]{article}%
\usepackage{amsmath}
\usepackage{graphicx}%
\usepackage{amsfonts}%
\usepackage{amssymb}

\begin{document}

\title{End of Several Quantum Mysteries }
\author{C. S. Unnikrishnan\\\textit{Gravitation Group, Tata Institute of Fundamental Research, }\\\textit{Homi Bhabha Road, Mumbai - 400 005}}
\date{{\small (First presented at the conference ``75 Years of Quantum
Entanglement'', Kolkata, January 6-10, 2011).}}
\maketitle

\begin{abstract}
I report on the discovery of quantum compatible local variables that are
shared between subsystems of quantum-conventionally entangled physical systems
such that they determine the correlations of spatially separated systems while
preserving strict Einstein locality. This puts an end to the mystery of spooky
action at a distance and alleged collapse at a distance, answering vital
questions, first raised in the EPR paper, on the behaviour of spatially
separated entangled systems. The solution helps to understand quantitative
measures of entanglement in a transparent way. It also provides new insight,
consistent with strict locality, of the physics of quantum teleportation and
related phenomena.

\medskip\ \ 

\end{abstract}

It is a widespread impression that the demonstrated violation of the Bell's
inequalities in experiments with spatially separated, quantum mechanically
entangled particles necessarily implies the violation of Einstein locality and
the existence of a nonlocal physical influence. This impression was reinforced
by the second generation experiments on Bell's inequalities that used time
variable polarization analyzers to ensure that the particles are outside each
other's light cone when the measurements are performed \cite{aspect}. However,
we emphasize at the outset that the violation of Einstein locality is implied
logically rigorously from Bell's work only when the theory one subscribes to
is a classical statistical theory called the hidden variable theory,
considered as a replacement for quantum theory, and not in the context of
quantum theory itself.

A local hidden variable theory or a local realistic theory is a
\emph{classical} statistical theory meant to replace quantum mechanics
\cite{belin}. John Bell's analysis of local hidden variable theories resulted
in the celebrated Bell's inequalities \cite{bell,bell-corr}. These represent
an upper limit on the correlation expected in such theories between results of
measurements on separated correlated quantum systems. In the standard
formulations, the magnitude of the correlation and its upper limit in a local
hidden variable theory are typically smaller than what are predicted in the
quantum mechanical description, for a wide range of settings of the
measurement apparatus. Thus, quantum mechanical correlations violate the
Bell's inequalities.

I have shown earlier that the quantum mechanical correlation function is a
unique consequence of some fundamental conservation law (that arising from a
space-time symmetry, like the conservation of angular momentum or from a
well-defined constraint associated with the physical problem), and therefore
all local hidden variable and local realistic theories are incompatible with
the fundamental conservation laws \cite{unni-x,unni-pram}. The basis of this
result was the discovery that the angular correlation function is simply an
average over the average angular momenta of one of the particles
\emph{conditional} on specific average values of the projection of angular
momentum ($+\hbar/2$ for example) for the other particle. Even the smallest
deviation, \emph{in either direction}, from the quantum correlation functions
in a loop-hole free experiment implies violation of conservation laws on the
average of the ensemble on which measurements are made. Hence, Bell's
inequality can be obeyed only by being incompatible with fundamental
conservation laws. This result shows that such theories are physically
inviable, and makes the demarcating criteria of Bell's inequalities their
experimental tests redundant \cite{unni-x,unni-pram}.

Another result of importance was that quantum correlations are a result of
pre-existing correlations at source, determined by the relevant conservation
laws and encoded \ as a relationship of relative phases of the subsystems
\cite{unnifpl}. The essential idea is that a classical conservation constraint
on any generalized momentum $p_{\mu},$ like the total angular momentum or
energy being a constant, directly translates to a constraint on the phase of
the combined quantum mechanical system since the quantum phase depends
linearly on these physical quantities (in the form $\frac{1}{\hbar}p_{\mu
}dx^{\mu}$ and appears in the complex exponential in the wave-function). Thus
a \emph{classical constraint} on a two-particle system of the form
$\overrightarrow{p}_{A}+\overrightarrow{p}_{B}=0$ will \emph{induce a
constraint on the quantum phases} of the two-particle wave-function. Any
attempt to encode the quantum correlations locally in the kinematical
variables themselves (like momentum, spin direction etc.) as has been done in
the context of local hidden variable theories is thus futile on first
principles. They have to be encoded in the relative phase as I had argued in
reference \cite{unnifpl}. This then implied that quantum correlations can
possibly be explained without any violation of Einstein locality, opening the
possibility of solving the fundamental quantum mystery associated with the EPR
problem and entangled systems. However, there were some points that required
clarifications and further development for a definite proof of validity of
strict Einstein locality \cite{desp}.

In this paper I provide that definite proof. I will show that the correlations
of the maximally entangled model system of the spin-1/2 singlet pair is
correctly implemented by strictly local variables that are shared at the
source. This is based on the discovery of \emph{quantum compatible local
variables} -- local quantities with positive classical probability
distributions that are compatible with both conservation laws as reflected in
the phases and also with the demonstrated superposition and wave-particle
duality for single particle systems, which is the core feature that generate
quantum phenomena. I will show how collapse at a distance is avoided by a
simple way of sharing prior classical conservation constraint in individual
random variables, similar to phases, with a definite relative coherence.

The proof is best presented in relation to Bell's own discussion on the
impossibility of finding shared variables that reproduce quantum correlations
in the spin-1/2 singlet case, thereby showing clearly how the Bell's theorem
is circumvented with the physically correct choice of shared variables.

Bell assumed that while measuring two valued discrete observable on quantum
systems like the two-particle spin singlet, the measurable values depend only
on the settings of the local measuring apparatus, and possibly on a set of
hidden variables. The maximally entangled state of two spin-1/2 particles, (or
polarization entangled photons) often employed in discussions and experiments
on Bell's inequalities is described by the wave function
\begin{equation}
\Psi_{S}=\frac{1}{\sqrt{2}}\{\left|  1,-1\right\rangle -\left|
-1,1\right\rangle \}
\end{equation}
where the state $\left|  1,-1\right\rangle $ is short form for $\left|
1\right\rangle _{1}\left|  -1\right\rangle _{2\mathrm{\ }}$, and represents a
definite value for spin projection of $+\hbar/2$ for the first particle and
$-\hbar/2$ for the second particle\emph{\ }if measured in any particular
direction. The state is a superposition of two such product states and it is
an entangled state. Two observers $A$ and $B$ make measurements on these
particles individually at space-like separated regions with time information
such that these results can be correlated later through a classical channel.
The local hidden variable description of the same system starts with the
functional restrictions on the outcomes $A$ and $B$ of measurements at the two
locations \cite{bell}.%

\begin{equation}
A(\mathbf{a,h})=\pm1,\quad B(\mathbf{b,h})=\pm1
\end{equation}
$A$ and $B$ denote the outcomes $+1$ or $-1$ of measurements $A$ and $B,$ and
$\mathbf{a}$ and $\mathbf{b}$ denote the settings of the analyzer or the
measurement apparatus for the first particle and the second particle
respectively. $\mathbf{h}$ are hidden variables associated with the outcomes.

The Bell correlation function is of the form
\begin{equation}
P_{B}(\mathbf{a},\mathbf{b})=\int d\mathbf{h}\rho(\mathbf{h})A(\mathbf{a,h}%
)B(\mathbf{b,h}),\mathrm{{where}\int d\mathbf{h}\rho(\mathbf{h})=1}%
\end{equation}
where $\rho(\mathbf{h})$ is the normalized statistical distribution of the
hidden variables.

This is an average over the product of the measurement results. $P(\mathbf{a}%
,\mathbf{b})$ is similar to the classical correlation function of the outcomes
defined by
\begin{equation}
P(\mathbf{a},\mathbf{b})=\frac{1}{N}\sum(A_{i}B_{i})
\end{equation}
The observed correlation is calculated using this formula, with observed
outcomes $A_{i}$ and $B_{i}.$

The quantum mechanical correlation for the same experiment is given by
\begin{equation}
P(\mathbf{a},\mathbf{b})_{QM}=-\mathbf{a}\cdot\mathbf{b}%
\end{equation}
(In terms of values of the projections of spin, the correlation function is
$P(\mathbf{a},\mathbf{b})_{QM}=-\mathbf{a}\cdot\mathbf{b~\hbar}^{2}/4$). The
essence of Bell's theorem is that the function $P(\mathbf{a},\mathbf{b})$ has
distinctly different dependences on the relative angle between the analyzers
for a local hidden variable description and for quantum mechanics.

In the local realistic model, $A$ (and $B$) can have simultaneous definite
values for various directions $\mathbf{a}$ (and $\mathbf{b}$) in the set
$\{+1,-1\}$, unlike the case in quantum theory where a definite value is
manifested only in a measurement for a particular direction without any way of
assigning values in the other unmeasured directions. Then the combination of
joint measurements
\begin{equation}
AB+A^{\prime}B-AB^{\prime}+A^{\prime}B^{\prime}=A(B-B^{\prime})+A^{\prime
}(B+B^{\prime})=\pm2
\end{equation}
because each observable takes values $\pm1,$ and the simultaneously assigned
values for $A$ and $A^{\prime}$ (or $B$ and $B^{\prime}$) can only be the
combinations $(+1,+1),$ $(+1,-1),$ $(-1,+1)$ and $(-1,-1).$ So the specific
combination of Bell correlation functions $P(\mathbf{a},\mathbf{b}%
)+P(\mathbf{a}^{\prime},\mathbf{b})-P(\mathbf{a},\mathbf{b}^{\prime
})+P(\mathbf{a}^{\prime},\mathbf{b}^{\prime})$ is an average of $\pm2$, and
lies between $+2$ and $-2.$ Its magnitude is bounded from above by $2.$ This
is the Bell's inequality. Looked at this way, it is clear that the inequality
arises from ignoring the fundamental premise of quantum mechanics --
superposition of states. Allowing the possibility of simultaneously assigning
definite values for the spin projection in two different and even orthogonal
directions for the same particle ($B=B^{\prime}=1$, for example, for 25\% of
the particles in this case) makes \emph{the subensemble violate quantum
superposition as well as the conservation constraint}. (We note that most
discussions of the EPR problem grossly misrepresents the meaning of local
realism, as meant by Einstein, and accuses him of wanting to ascribe
simultaneous values to quantum mechanically incompatible physical quantities.
This is incorrect, as evident from his own words quoted in appendix A. All he
said was that the `physical state' of the distant particle should be described
by a unique wave-function with one-to-one correspondence, and since the
physical state could not be changed at a distance, the wave-function, which
was supposed to be its complete representation, also should not change by some
measurement done on part of the system far away. However, since the
wave-function does depend on what kind of measurement is done on a subsystem,
there is no one-to-one correspondence between the physical state and the
wave-function, hence quantum mechanics is incomplete. This was the EPR
argument. \emph{There is no implication or desire that, for example, }%
$x$\emph{ and }$p_{x}$\emph{ or \ }$S_{x}$\emph{ and }$S_{y}$\emph{ values
exist simultaneously for a quantum}. Simple wave-particle duality that
Einstein pioneered and supported prohibits that. The point was that QM could
not ascribe a wave-function to the distant particle of the entangled pair
before one measurement, and after the measurement the wave-function depends on
the kind of measurement chosen even though the physical state itself could not
have changed at a distance, including the emergence of a physical state when
none uniquely existed according to QM - see appendix A for clarity on this).

Since $A(\mathbf{a})=-B(\mathbf{a}),$ a perfect anti-correlation for the
singlet pair when the measurement setting are the same at $A$ and $B,$ Bell
replaced expressions of the form $P_{B}(\mathbf{a},\mathbf{b})=\int
d\mathbf{h}\rho(\mathbf{h})A(\mathbf{a,h})B(\mathbf{b,h})$ \ with
$P_{B}(\mathbf{a},\mathbf{b})=-\int d\mathbf{h}\rho(\mathbf{h})A(\mathbf{a,h}%
)A(\mathbf{b,h}).$ This obviously takes a drastic deviation from the
originally announced program of \emph{adding more variables to quantum
mechanics} to make it more complete since this step of assigning simultaneous
definite values to quantum mechanically incompatible observables is
inconsistent with the core idea in quantum mechanics -- \ that of
superposition. For the case of spins, for example, this step allows ascribing
definite values a priori, ($+1,+1$) for example, for the Z and X directions.
We can show how this step is inconsistent with even single particle quantum
mechanics involving interference \cite{unni-real-single}. It is this step that
violates the possibility of superposition that leads to the inequality. This
is discussed elsewhere \cite{unni-spie}. Right now we proceed with the proof
of Einstein locality.

We assume, as in conventional hidden variable theories, that there is some
shared generalized vector variable associated with the particles that separate
into relativistically spatially separated regions after an interaction at a
common `source region'. We denote this random unit vector $\hat{\lambda}.$ For
example, with one particle of the singlet we can associate $\hat{\lambda}$ and
with the other $-\hat{\lambda}.$ The vector $\hat{\lambda}$ is uniformly
distributed over the 2D sphere with $\rho(\hat{\lambda})\geq0$. Let the
measurements on the particles are done independently in spatial directions
$\hat{a}$ and $\hat{b}.$ In the discussion by Bell \cite{bell-corr}, it was
shown that such $\hat{\lambda}$, from a set of ordinary vectors, could not
reproduce the two-particle correlations correctly even though single particle
measurement results of $A(\mathbf{a})=\pm1$ with 50\% probability each as well
as $A(\mathbf{a})=-B(\mathbf{a})$ for every pair can be reproduced. A
prescription chosen normally in such demonstrations is $A(\mathbf{a}%
)=sign(\hat{\lambda}\cdot\hat{a}).$ Note that this is deterministic, given the
direction of $\hat{\lambda}.$ The EPR argument does not discuss lack of
determinism as the deficiency of quantum theory, and answering the EPR query
does not demand making single particle quantum behaviour deterministic. In
fact, the central point there is how one could get perfect determinism for
correlations in certain directions while the individual measurements are
local, independent and probabilistic. Our earlier remarks on conservation laws
and phases should make it clear why this prescription does not work and why
this should not be the starting point in any case. \emph{What is needed is a
shared variable that is compatible with wave-particle duality (superposition
in quantum mechanical terms)}, and that is the key point. Anything else is
naive and anti-quantum mechanical. Total determinism and realism is
incompatible with even single particle quantum mechanics and its core
characteristic of superposition. Multiple response to a definite state is a
direct consequence of superposition, both in quantum and classical physics.
The central issue in the EPR problem is locality and not determinism.

There is a solution that is compatible with quantum superposition as well as
Einstein locality. We note that if $\left(  \hat{\lambda}\cdot\hat{a}\right)
^{2}=1,$ then $\hat{\lambda}\cdot\hat{a}=\pm1$ and the measurement results
$+1$ or $-1$ are determined by whether $\hat{\lambda}\cdot\hat{a}$ is $+1$ or
$-1,$ up to some phase (`parallel' or `anti-parallel'). So, $A(\mathbf{a})$
and $B(\mathbf{a})$ are $\hat{\lambda}\cdot\hat{a}$ and $\hat{\lambda}%
\cdot\hat{b}.$ Then the correlation is an average over the random vector
$\hat{\lambda},$
\begin{align}
P(\hat{a},\hat{b})  &  =\left\langle \left(  \hat{\lambda}\cdot\hat{a}\right)
\left(  -\hat{\lambda}\cdot\hat{b}\right)  \right\rangle _{\hat{\lambda}%
}\nonumber\\
&  =-\left\langle \left(  \lambda_{1}a_{1}+\lambda_{2}a_{2}+\lambda_{3}%
a_{3}\right)  \left(  \lambda_{1}b_{1}+\lambda_{2}b_{2}+\lambda_{3}%
b_{3}\right)  \right\rangle _{\hat{\lambda}}%
\end{align}
If the vector $\hat{\lambda}$ is such that $\lambda_{1}^{2}=\lambda_{2}%
^{2}=\lambda_{3}^{2}=1,$ then%
\begin{equation}
P(\hat{a},\hat{b})=-\left\langle
\begin{array}
[c]{c}%
\hat{a}\cdot\hat{b}+\lambda_{1}\lambda_{2}a_{1}b_{2}+\lambda_{2}\lambda
_{1}a_{2}b_{1}\\
+\lambda_{2}\lambda_{3}a_{2}b_{3}+\lambda_{3}\lambda_{2}a_{3}b_{2}+\lambda
_{3}\lambda_{1}a_{3}b_{1}+\lambda_{1}\lambda_{3}a_{1}b_{3}%
\end{array}
\right\rangle _{\hat{\lambda}}%
\end{equation}
It is clear that this will give the average correlation agreeing with quantum
mechanical correlation when averaged over the random shared vector. However,
\emph{we also need perfect correlation}, \emph{without any averaging when the
directions }$\hat{a}$\emph{ and }$\hat{b}$\emph{ match}. This is easily
achieved by setting the \emph{algebra of the local components} of the shared
vector as
\begin{align}
\lambda_{1}\lambda_{2}  &  =-\lambda_{2}\lambda_{1}=i\lambda_{3}\nonumber\\
\lambda_{2}\lambda_{3}  &  =-\lambda_{3}\lambda_{2}=i\lambda_{1}\nonumber\\
\lambda_{3}\lambda_{1}  &  =-\lambda_{1}\lambda_{3}=i\lambda_{2}%
\end{align}
Then we get
\begin{align}
P(\hat{a},\hat{b})  &  =-\left\langle \hat{a}\cdot\hat{b}+i\hat{\lambda}%
\cdot\left(  \hat{a}\times\hat{b}\right)  \right\rangle _{\hat{\lambda}}\\
&  =-\cos\theta-\left\langle i\hat{\lambda}\cdot\hat{n}\right\rangle
_{\hat{\lambda}}\sin\theta
\end{align}
with the well known identity inside the averaging brackets. $\hat{n}$ is the
unit vector in the direction of $\hat{a}\times\hat{b}.$ When $\hat{a}$ and
$\hat{b}$ match, $\hat{a}\times\hat{b}=0,$ and guarantees perfect correlation
$-1$ for \emph{every} measurement on the pair of the singlet. Since
$\hat{\lambda}$ is a random vector, averaging makes the $\hat{\lambda}%
\cdot\left(  \hat{a}\times\hat{b}\right)  $ term zero for an ensemble. Thus,
by choosing a particular \emph{local algebra} for the components of the shared
hidden vector we have circumvented the Bell's theorem and reproduced the
correlations seen in experiments. Another view on this is to say that the
correlation is an average over a pure relative phase containing a random
vector, $\exp(i\hat{\lambda}\cdot\hat{n}\,\theta)$ where $\hat{n}\sin
\theta=\hat{a}\times\hat{b}$ and $\theta=\cos^{-1}(\hat{a}\cdot\hat{b}).$ This
filters out just the imaginary part of the pure phase, much in the line
suggested in our earlier work \cite{unnifpl} on this issue. I assert that this
is exactly how nature implements the correlations with perfect adherence to
Einstein locality. We have discovered \emph{local variables that are quantum
compatible} that allow coding the shared information without violating
Einstein locality or the requirements arising from quantum superposition.
There is absolutely no spooky action at a distance nor is any nonlocal
collapse of the state at a distance. The simple algebra of the components of
the shared vector guarantees both two-valued measurements as well as the
correct correlations. The EPR-Bohm problem is completely resolved respecting
perfect Einstein locality.

For photons, it is not the same set of $\hat{\lambda},$ responsible for the
correlation of spin-1/2 particles, that will share the correlation information
at source even though the polarization is a two-component physical quantity.
Since a rotation of angle $\pi/2$ radian rather than $\pi$ takes the state to
its orthogonal conservation pair, angles are multiplied by 2 when the photon
problem is mapped to the spin-1/2 problem. In other words, when ideas of
angular momentum conservation is used for `singlet' created with photon
polarization, the angles in the cosine function has to be doubled because the
angle space is doubly covered.

Our resolution answers the EPR query whether the quantum mechanical
description could be considered complete. The local variable that is shared
between the subsystem does contain the information of all possible
measurements, without determinism -- these variables are not fixed directions
specified by real vectors with 3 real numbers in 3D space as Bell considered,
in the case of spin variables, but a set of numbers or a matrix for each
spatial dimension. The local algebra of these components are anti-commuting,
and therefore compatible with superposition for spin components. When a
measurement is done on one sub-system neither the physical state nor its
associated shared variable changes for the spatially separated partner. By the
very nature of these local variables, they are oppositely correlated for the
same setting, and also reproduces the correct angular dependence when averaged
over the ensemble of measurements.

It may be noted that the same scenario correctly describes usual cases of
single particle measurements involving spin or polarization, for unpolarized
or polarized particles, with an appropriate distribution for the generalized
quantum compatible vectors $\hat{\lambda}.$ It can also correctly describe
mixed state correlations in the multi-particle case.

In conclusion, quantum mysteries that have been in discussion for decades
concerning how correlations of measurements in the microscopic world are
maintained over spatially separated regions respecting Einstein locality,
without any spooky action at a distance, are resolved. We have shown by
explicit construction that Einstein locality is strictly valid in nature,
ending a 75-year old universal worry \cite{kolkata}. Entanglement is not to be
equated to nonlocality. \bigskip

\noindent{\large Appendix A}{\Large \medskip}

Excerpts from Einstein's letter to K. Popper (reproduced in `Logic of
Scientific Discovery') explaining his view that the wave-function description
is incomplete:

``Should we regard the wave-function whose time dependent changes are,
according to Shrodinger equation, deterministic, as a complete description of
physical reality, and should we therefore regard the (insufficiently known)
interference with the system from without as alone responsible for the fact
that our predictions have a merely statistical character?

The answer at which we arrive is the wave-function should not be regarded as a
complete description of the physical state of the system.

We consider a composite system, consisting of the partial systems A and B
which interact for a short time only.

We assume that we know the wave-function of the composite system before the
interaction -- a collision of two free particles, for example -- has taken
place. Then Schrodinger equation will give us the wave-function for the
composite system after the interaction.

Assume that now (after the interaction) an optimal measurement is carried out
upon the partial system A, which may be done in various ways, however
depending on the variables which one wants to measure precisely -- for
example, the momentum or the position coordinate. Quantum mechanics will then
give us the wave-function for the partial system B, and it will give us
various wave-functions that differ, according to the kind of measurement which
we have chosen to carry out upon A.

Now it is unreasonable to assume that the physical state of B may depend upon
some measurement carried out upon a system A which by now is separated from B
(so that it no longer interacts with B); and this means that the two different
wave-functions belong to one and the same physical state of B. Since a
complete description of a physical state must necessarily be an unambiguous
description (apart from superficialities such as units, choice of the
coordinates etc.) it is therefore not possible to regard the wave-function as
the complete description of the state of the system.''

\smallskip

\noindent\textbf{Comment}: By using the quantum compatible shared variables,
we have managed to incorporate in a strictly local theory `two different
wave-functions belonging to one and the same physical state of B' without
conflict, answering the query implied in the EPR paper of how the
multiparticle correlation can be deterministic while the individual
measurements are local, independent and probabilistic.

\end{document}